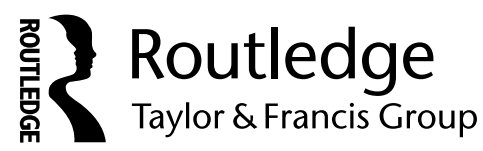

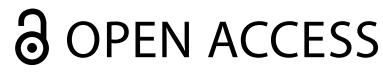
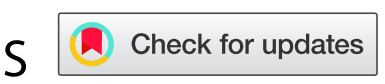

# A new digital divide? Coder worldviews, the 'Slop economy,' and democracy in the age of AI

Jason Miklian[a] and Kristian Hoelscher[b]

[a]Centre for Global Sustainability, University of Oslo, Oslo, Norway; [b]Peace Research Institute Oslo, Oslo, Norway

**ABSTRACT**

Digital technologies are transforming democratic life in conflicting ways. This article bridges two perspectives to unpack these tensions. First, we present an original survey of software developers in Silicon Valley, interrogating how coders' worldviews, ethics, and workplace cultures shape the democratic potential and social impact of the technologies they build. Results indicate that while most developers recognize the power of their products to influence civil liberties and political discourse, they often face ethical dilemmas and top-down pressures that can lead to design choices undermining democratic ideals. Second, we critically investigate these findings in the context of an emerging 'new digital divide', not of internet access but of information quality. We interrogate the survey findings in the context of the 'slop economy', in which billions of users unable to pay for high-quality content experience an internet dominated by low-quality, AI-generated ad-driven content. We find a reinforcing cycle between tech creator beliefs and the digital ecosystems they spawn. We discuss implications for democratic governance, arguing for more ethically informed design and policy interventions to help bridge the digital divide to ensure that technological innovation supports rather than subverts democratic values in the next chapter of the digital age.

## Introduction and knowledge base

Scholars are increasingly concerned with how digital technologies are reshaping democracy (e.g., Bernholz et al., 2021; Fuchs, 2022). Digital media and internet tools can empower citizens, broaden political participation, and disseminate information at unprecedented scale. However, these same tools can erode trust in institutions, amplify misinformation, deepen polarization, and facilitate authoritarian practices. How and when technology serves as a tool for democratic deepening versus a vehicle for democratic backsliding remains unclear (Cupać et al., 2024; Miller and Vaccari 2020).

**CONTACT** Jason Miklian jason.miklian@globe.uio.no Centre for Global Sustainability, University of Oslo, Sandakerveien 130, 1121, Oslo, Norway

Acknowledging that human choices in technology design can influence democratic outcomes, two critical questions emerge. First, how might the worldviews, ethics, and beliefs of technology developers become embedded in digital products that influence civic life? Second, how do their products collectively shape information access and quality in a democratic society? These questions motivate our study. We argue that understanding the democratic impacts of technology requires examining both the production-level (the beliefs and ethics of technology creators) and the consumer-level (the structure of the digital ecosystem that users inhabit) of analysis.

A rich tradition of technology research suggests that the assumptions, values, and politics of design communities shape what technologies get built and how they function (Jungherr & Rauchfleisch, 2025). Therefore, political and ethical orientations of software developers and tech company leaders could influence technologies in ways that affect civic participation, rights, and discourse.[1] Digital product design choices can produce profound societal consequences, and those choices may trace back to the beliefs of the humans behind the screen.

While there are some broad assumptions about tech worker beliefs,[2] there is limited systematic knowledge about what coders and developers believe about democracy and society, either as a snapshot or across time. Tech industry workers' have traditionally aligned with broadly liberal democratic social values, but in ways that emphasize technical expertise and/or techno-managerial solutions to collective challenges. This is grounded in the 'Californian Ideology', a prototypical Silicon Valley socio-political worldview characterized by a blend of libertarianism, neoliberalism, individualism, technological primacy, and countercultural beliefs (Barbrook & Cameron, 1996). Recent evidence suggests US tech workers remain socially liberal and egalitarian, but also anti-establishment (Cook, 2020; Daub, 2020). Others suggest a rightward shift in the politics of Silicon Valley (Golumbia, 2024), revealing how the industry is becoming increasingly diverse in their socio-political beliefs and worldviews. American tech workers exhibit a unique blend of liberal and anti-establishment leanings (Massanari, 2024; Selling & Strimling, 2023), complicating stereotypes about Silicon Valley's political monoculture. From a more contemporary perspective, corporate leaders wield significant influence in directing tech innovation, and their ideological stances can steer platform policies (Brockmann et al., 2021). We also know regulators and investors shape incentive structures (Globig et al., 2023). Yet rank-and-file developers, those writing the code, are often treated as apolitical actors simply implementing technical requirements.

Our study explores this omission by asking coders themselves what they believe about the relationship between technology and democracy, if their personal ethics and ideologies influence their work, and if they observe political biases in their company and industry. This inquiry aligns with calls to unite sociopolitical and design perspectives in tech research, and extends recent investigations into tech worker culture and politics (Mitra, 2024; Tan et al., 2023). Our research builds on such insights by asking Silicon Valley coders about democracy and social impact – to our knowledge, the first survey of its kind. We posit that coder beliefs and ethics are an under-appreciated 'upstream' factor in how technology impacts democracy downstream.

This is also of interest as rank-and-file tech worker worldviews may differ from leaderships. Indeed we may expect a negotiation of competing values within large organizations (and within individuals themselves on such issues), and there is a rich literature highlighting the moral, ethical and political tensions between employees and their bosses in field such as media, politics and technology (Bietti, 2021; Di & Nishikawa, 2025; Fan, 2019; Kucinskas & Zylan, 2023). While some coders work in environments where a CEO steers their politics (as with Twitter/X since 2022), tech employees don't generally work for firms with an overt political agenda. Instead, they seek to create venues for *engaging content*. Coders (especially for public-facing firms) thus mostly aren't faced with forcing a specific political viewpoint, but about supporting the agenda of 'just asking questions'-style content creation to farm engagement (Lindman et al., 2022; Monsees et al., 2023). Many technology firms have found that material that raises emotions is great for metrics, especially material that exploits both sides of political polarization (Saura García, 2024; Unver, 2017).

This recognition intersects with a new global digital dynamic: as internet *access* spreads globally, the *quality* of digital content diverges sharply. The 'digital divide' of past decades referred to unequal access to the internet across societies. That gap is closing – 5.5 billion people use the internet as of early 2025, and by 2030 nearly 90% of people age 6 and up will be online (Statista, 2025). However, the proliferation of access has revealed a new divide: one based not on *whether* one is online, but *what kind of internet* one can access. This is largely determined by an individual's ability to buy themselves away from informational platforms where they are often the product themselves.

We posit that global society is bifurcating into two broadly parallel information environments; digital elites and digital commoners. We categorize these modes of digital information consumption as an illustrative heuristic to differentiate 'high' and 'low' value digital content and how these may differ. Of course, individual use and consumption of digital information exists on a continuum between these two types of information environments rather than an either/or placement. We position the argument as a binary to highlight what a 'typical' user in both camps experiences.

On one side are the 'digital elites' – those with the means, skills, or institutional support to obtain high-quality information and online experiences. This group enjoys reliable news sources, can afford ad-free subscriptions or premium content, and benefits from platforms and regulations that attempt to uphold standards of accuracy, privacy, and democratic values. Their internet experience includes credible journalism (e.g., The New York Times, BBC), fact-checked content, and fewer mis/disinformation traps.

On the other side are the billions of 'digital commoners' whose online experience is dominated by what Read (2024) coined the 'slop economy'. This concept encapsulates this grim reality of the second-tier internet, a 'thriving, global gray-market' of Artificial Intelligence (AI) content mills and algorithmic spam that clogs search results with nonsense and pollutes their information ecosystem. Platforms optimized for advertising revenue flood users' feeds with maximal content for minimal cost, which increasingly means auto-generated text and videos. Already, an estimated 40% of content consumed is churned out by AI, recombining material scraped from human authors into a synthetic slurry of words (Spennemann, 2025).

In the slop economy, users subsist on a near-zero-cost diet of content that is abundant but of limited 'factual nutrition' – meaning clickbait blogs, sensational headlines, and algorithmically generated text and videos of dubious accuracy (Klee, 2024; Koebler, 2024). These users cannot afford (or forego) paywalled content and rely on free or mostly-free services that are funded by ads or data harvesting. As a result, their information ecosystem is rife with spammy, AI-generated content (Read, 2024; see Sarkar (2025) for a rebuttal). Such slop ranges from deceptive AI-written news articles with fake bylines, to autogenerated non-factual how-to guides, to political memes and clickbait optimized for engagement (Klincewicz et al., 2025). This content is not always malicious *per se*; much of it is simply hallucinatory text and images that bend reality altogether.

Why does this divide matter for democracy? Democracies depend on an informed citizenry and a shared base of facts for deliberation (e.g., Carpini & Keeter, 1997; Milner, 2002). If a large segment of the population is consuming fundamentally inferior information – essentially digital junk food – their capacity to participate meaningfully in democratic life may be compromised (Chambers, 2021; Coeckelbergh, 2024; Dodds et al., 2025). In countries without strong public service media to provide universal high-quality information, this divide runs along economic lines, as some pay for access to high-quality information, leaving those unable to pay with the noisy, automated, and contested free digital information environment (Hamilton & Morgan, 2018; Moore, 2024). Such stratification in knowledge environments could mean that socio-economic status determines one's ability to make informed political decisions. Citizens fed on false or incoherent narratives may become susceptible to extremism or disillusionment, *even without targeted disinformation campaigns*, simply through the low quality content in their information feeds.

Examples abound. In many low-income or marginalized communities, internet users' primary portals are free apps pre-loaded on cheap smartphones or accessed via zero-rated mobile plans. These apps – news aggregators, entertainment platforms, or social media clones – serve content with minimal curation. For instance, Facebook's Free Basics program offered a slice of the internet free of charge, but only a limited set of sites hand-picked by Facebook with superficial content and advertising in a tightly walled experience. Although ultra-cheap data plans have made the internet accessible worldwide, the practical reality is that paywalls fence off high-quality content. A user who can barely afford a $3 monthly 1GB data pack is unlikely to spend extra on a news subscription. In the United States, a wealthier individual might pay for several reliable news subscriptions, whereas a poorer individual may access free news on X, TikTok or YouTube, where misinformation and clickbait proliferate.

Despite these trends, scholarly research on this new quality-based digital divide is scant (Kassa, 2024). Most existing studies of digital inequality still focus on access or digital skills, not on content disparities (Raihan et al., 2024). Similarly, the social impacts of algorithmic content are often studied in the context of social media or specific misinformation incidents, rather than as a broad information ecosystem issue. But the slop economy may have more widespread influence on everyday political knowledge than the much-studied phenomena of electoral disinformation. In other words, while scholars scrutinize how X or Meta might sway voters (Boulianne, 2020; Spierings & Jacobs, 2014) or shape political engagement (Chan & Yi, 2024), a parallel universe of lesser-known apps and aggregate sites is quietly shaping (or misshaping) the civic consciousness of a huge portion of the globe.

This article brings together coder beliefs and the digital divide to explore how the *ideological design of technology* and the *unequal consumption of technology* influence democracy. We posit that these dynamics are interconnected. If, for example, tech company leaders prioritize engagement metrics and ad revenue over safeguards for truth and civility, they will construct a slop-rich environment. Conversely, if developers had greater ethical autonomy and a commitment to democratic values, they might design products that resist 'slopification'. By examining both developers and users, we aim to paint a fuller picture of the democratic dilemmas of digital innovation with new empirical evidence and conceptual insights at the intersection of technology design and democratic development.

The article proceeds as follows. In Materials and Methods, we describe our original survey of Silicon Valley software developers. In Results we report key findings, including developers' perceptions of technology's influence on democracy, ethical challenges they face, and the role of corporate leadership ideologies. In the Discussion, we present insights on the new digital divide, highlighting how low-quality information platforms shape political knowledge, arguing that coder beliefs and the slop economy require connected consideration. Finally, we offer concluding thoughts on bridging gaps for further research on digital development-democracy interactions.

## Materials and methods

We employed a survey of software developers to capture their beliefs and experiences related to technology's social impact. We designed an original survey of coders, software engineers, and related roles to systematically map their worldviews about technology and democracy, as per the De Leeuw et al. (2012) methodological guidance on surveys and internet-based surveys. The survey was conducted via smartphone in mid-2024 by the firm RIWI in Silicon Valley and the broader Bay Area in California.[3]

We obtained 421 complete responses. 60% of the sample reported working in large tech firms (defined as >1000 employees), 25% in mid-size companies, and 15% in small startups (<50 employees). The sample was predominantly male (≈68%) and aged 25–44 (median age≈30), reflecting industry demographics (SVIRS, 2025). In terms of self-described political leanings, ≈52% identified as left-leaning or liberal, 22% as centrist/moderate, 14% as right-leaning or conservative, and 12% chose not to disclose. We treat the data as an exploratory descriptive mapping of coder beliefs. Responses were aggregated and analyzed using descriptive statistics (frequencies, cross-tabulations by subgroups like firm size or political ID) and identification of notable patterns.

Most existing developer surveys (e.g., industry surveys by Stack Overflow or HackerRank) seldom ask about social impact or democratic values, concentrating instead on technical skills and job preferences. Our survey thus centered on questions of ethics, politics, and perceived social impacts of the respondent's work (see Appendix 1). We developed the questionnaire through an iterative process, incorporating feedback from social scientists and software practitioners for clarity and relevance. Key themes included:

### *Perceptions of technology's influence*

To what extent do developers believe the products they create affect societal outcomes like democratic governance, elections, or civil liberties? For example, we asked for level of agreement with statements such as '*Technology products can unintentionally undermine democratic ideals (e.g., free elections, civil liberties)*' and '*Products at my company have influenced democracy (positively or negatively)*'.

### *Workplace ethical climate*

Whether and how often developers face ethical dilemmas or pressures at work. Sample question: '*Have you ever felt pressured by your employer to implement a feature in a way that might restrict human rights or freedoms (e.g., aiding surveillance)?*' and if so, how they responded. We also inquired if they ever *regretted* the social impact of a project they contributed to.

### *Leadership and ideology*

The role of leadership's worldview or founder ideologies in shaping product design. We asked if respondents felt that personal beliefs of top executives influenced their builds. We also queried respondents' own political leanings to analyze whether ideological orientation correlates with their views on tech's role in society.

### *Democracy-focused tech vs. general tech*

We differentiated respondents who self-identified as working on 'civic tech' or tech explicitly intended to support democratic governance (e-voting platforms, government transparency tools, etc.) versus general consumer or enterprise tech. This allowed us to compare perspectives between those in explicitly democracy-relevant projects and those in mainstream tech.

## Results

Five key themes emerged from our analysis of the data. First, *coders widely acknowledge that technology can shape democratic outcomes, and many worry about unintentional harms*. Second, *developers see leadership's ideology as a strong influence on product decisions*. Third, *political partisanship among coders is less stark than expected, with a general consensus about certain risks.* Fourth, *ethical dilemmas are common, and many developers have felt pressure to act against their social values.* Fifth, *even those creating 'pro-democracy' tech feel their products may undermine democracy*.

### *Technology's influence on democracy is recognized – and often feared*

A large majority of developers in our sample believed that the products and platforms tech workers create influence democratic principles and processes. Asked whether 'technology creators can deliver products that unintentionally undermine democratic ideals (like civil liberties or fair elections),' 78% of respondents agreed or strongly agreed. About half of this group saw unintended negative impacts in their own company's products. This demonstrates a noteworthy level of awareness: most coders are not blindly techno-optimistic or assume their creations are neutral. Rather, even well-intentioned tools can erode democratic values. Our data show that rank-and-file developers largely accept this critique: only 12% in our survey disagreed that tech products can unintentionally harm democracy.

However, many developers still believe in tech's positive potential for democracy, indicating a nuanced rather than simply pessimistic position. 64% agreed that 'digital technology has the potential to strengthen democratic participation.' aligning with research on technology and citizen empowerment (Shin et al., 2024). The prevailing sentiment was *ambivalent power*: tech is a powerful tool that may bolster or undermine democracy, depending on how it's used. Notably, only a small minority (~10%) took an extreme view either way (purely utopian or purely dystopian about tech and democracy). Most see a double-edged sword, whereby decisions in the development process steer the direction it is swung.

### CEO and leadership ideologies strongly shape product outcomes

Developers overwhelmingly reported that the worldview of company leadership influences what they build and how they build it. When asked, 'How much do you feel the CEO/founder's personal beliefs and values influence the design or features of your products?', 81% answered 'a great deal' or 'a fair amount.' This was consistent across company sizes (slightly more pronounced in large firms (85%), versus 78% in startups). Our findings suggest the same holds in tech product design. If most coders feel they are implementing their CEO's worldview, this can concentrate socio-political power in the hands of executives.

Developers at small startups were the most likely to report negative social impacts of their products: 38% said they believe their product is currently having a net negative impact on society (versus 11% in large firms). This could be interpreted in several ways. Small start-ups may be working on riskier, untested ideas where social harms are more plausible, or they may lack resources and oversight to mitigate harms. It could also reflect a difference in candor or culture; perhaps big firm employees are more cautious about admitting harm. Regardless, it challenges a common assumption that there is something inherently more harmful about Big Tech actors; developers for startups are at least as likely to acknowledge negative externalities.

### Partisan divides among developers are muted

Self-described liberals were more likely to emphasize tech's harms (e.g., agreement that tech can undermine democracy), whereas self-described conservatives were slightly more skeptical of tech's benefits (some expressed concern that platforms censor voices). Yet on many core questions, developers across the political spectrum shared similar concerns. For instance, 79% of liberals and 74% of conservatives agreed that CEO worldviews influence products. Both groups reported high rates of ethical dilemmas (~80%, see next section). One notable split was that centrist or politically unaligned developers were the most likely to be indifferent about these issues. Over half of self-identified centrists chose 'Neutral/Not sure' on whether tech influences democracy, compared to under 20% of those with a partisan leaning. This suggests that tech 'moderates' are the least worried by ethics debates (see Limitations for more).

In contrast, those who identified at either end (left or right) were more aligned, possibly because both far-left and far-right tech workers harbor anti-establishment sentiments that make them critical of Big Tech's societal effects. The usual left-right political divide may defer to a more general technologist perspective (Selling & Strimling, 2023). Thus, efforts to improve tech's democratic outcomes might require different approaches than what might work in other settings, as the narrative of 'Democrats vs. Republicans' doesn't map cleanly onto the tech creator context. Similarly, we found no particular gender differences in responses, which is notable given the 2:1 male bias in tech workers (also reflected in our survey). The fact that political identification and gender were not axes of differentiation among respondents may suggest that there is some 'core set of beliefs' among tech workers that transcend typical differentiators that future research should examine further, as in the large foreign-born cohort amongst tech workers in Silicon Valley more generally (c. 40% of workers)

### *Ethical dilemmas and pressure are common*

Ethical challenges appear to be ubiquitous. Over half of our sample agreed to: 'Have you ever felt regret about the social impact of a project you worked on at your current company?' We next asked: 'If you felt pressure to restrict certain human freedoms or liberties in your work (for example, being asked to create a security feature that could be used to surveil citizens), what would you do?' 74% said they *would still implement it*, while 26% would refuse or escalate the issue. This suggests that although many developers report to have strong personal ethics, the majority still comply with problematic directives, possibly due to job pressure, a sense of duty, or belief that someone else will do it if they don't.

In other words, resistance is hard; refusing a task could be career-limiting. This speaks to the oft-cited tension between employee values and organizational demands in tech companies. Recent years have seen tech employee protests, such as Google employees opposing a Pentagon AI contract in 2018 (Shane & Wakabayashi, 2018), or recent Microsoft protest over ties with the Israeli military (Buncombe, 2025). Such incidents indicate that ethical conflicts within tech companies are common. Our data suggests these conflicts are not rare one-offs but rather a fairly routine part of being a developer. In fact, about 80% of respondents agreed that 'software developers face ethical issues in their work.'

Developers in certain sectors report stronger ethical pressures. Those working for social media or advertising technology firms reported the highest willingness to implement potentially harmful features, and the highest observation of ideological bias in their companies. Nearly 90% of surveyed social media employees observed political or ideological biases being 'injected' into algorithms or content at their firm. They also answered the ethical compliance question with 'I would do it without significant objection,' at a higher rate than others (over 50% agreed). Respondents in enterprise software or education sectors were more likely to say they would resist similar pressures. Our sample suggests that those working on social media products are keenly aware of bias and often push forward regardless, perhaps rationalizing that the benefits or the inevitability of the task outweigh their discomfort.

### *Democracy-oriented technologies don't inoculate against negative impacts*

15% of respondents stated that they work on products explicitly aimed at supporting democratic ideals like online voting systems, petition platforms, fact-checking tools, or civic engagement apps. Yet this group was among the most critical about democratic outcomes. 68% of those working on 'democracy tech' said they have seen their product unintentionally undermine the very democratic ideals it was meant to support. These findings underscore the complexity of designing for democracy: even when seeking to improve democratic outcomes, mis-execution or misuse can undermine such goals. Notably, this group did not report any less pressure or ethical strain.

Our survey suggests a community of tech creators that is self-aware, concerned, yet often constrained. Developers widely believe their work has political and social ramifications; many harbor ethical reservations about what they've done or might be asked to do. It also implies that developers and coders navigate a series of overlapping ethical and value-oriented dilemmas in their positions, wherein the politics of beliefs and employment make binary decisions about 'is what I am building good for democracy?' More multi-faceted and complicated than is oten assumed. They also point to organizational culture and leadership values as important determinants of whether their products align with democratic values. While they largely share concerns regardless of personal politics, the decision to act ethically (whistleblow, refuse tasks) is rare, implying the need for stronger support systems.

## Discussion

Next, we synthesize these findings to explore their scholarly implication. We organize the discussion around three themes: (1) **Democracy by (Tech) Design:** how developer beliefs and ethics can influence democratic outcomes through design choices but may clash with leadership demands; (2) **Democracy in the Digital Public Sphere:** how the slop economy and information-quality divide are impacting democratic participation; and (3) **Bridging the Gaps:** integrating the micro and macro perspective to empower responsible tech creation while improving the digital ecosystem for end users.

### *Democracy by (Tech) design*

One clear message from our survey is that human agency within technology companies matters for the democratic character of tech products. Developers are not mere cogs executing neutral instructions; their values, their leaders' ideologies, and the internal culture shapes product features that impact society. Developers recognize that CEO worldviews influence products, and are ethically conflicted. This finding reinforces that *who* designs technology (and under what belief system) matters for its democratic impacts. We might think of coders and engineers as an emergent stakeholder class in democratic governance, given their role in constructing tools that mediate civic life. As Davis (2020) notes, the cultural and political beliefs of tech communities imprint onto what they produce. Our data confirm that imprinting is happening, sometimes intentionally and sometimes through unspoken biases.

The findings support the importance of building in 'democracy by design': incorporating democratic principles of fairness, transparency, accountability, and inclusion into the design process. The need for proactive design for social impact is not new (Collingridge, 1980), but our findings show it is still not standard practice. Most coders felt their companies address social impact only after issues become public, which aligns with the reactive pattern companies employ in dealing with, e.g., content moderation scandals (Rim & Ferguson, 2017). A shift to anticipatory governance of technology (Nemitz, 2018) would empower developers to voice concerns early and design in mitigatory features. This is also not a new idea, as science, technology and society (STS) studies scholars have built similar responsible innovation models that incorporate ethics and values into innovation (Grunwald, 2011; Tatum, 2004). Some companies have ethics review boards or 'red team' exercises for AI products. Yet, if the core incentive structure driven by leadership ideology or profit remains unchanged, such measures can be perfunctory and few managers will prioritize an ethical concern that could slow growth.

Our survey also indicated that a large portion of developers may welcome clearer ethical guidelines from leadership. Many developers want to build responsibly but need top-down support to do so without fear of reprisal. In practical terms, tech companies might institutionalize a 'democracy check' impact assessment whereby features are evaluated for how they might affect election integrity, public discourse, or privacy rights. It should carry weight in go/no-go decisions, similar to how data privacy reviews have clout post-GDPR. While it is unrealistic to expect profit-driven companies to become guardians of democracy, our research suggests a large internal constituency will support deeper changes if empowered. Therefore, strengthening channels for internal advocacy could amplify coder voices for democratic design principles.

However, we must also confront the reality that many developers will comply with problematic orders due to job security or inertia, and blaming workers for implementing anti-democratic company policies from the C-suite or ownership can become a means for leadership to skirt responsibility. Industry-wide standards or codes of ethics similar to medical or legal professions could help establish a baseline of acceptable practice and offer worker protections. There have been attempts (e.g., ACM Code of Ethics), but enforcement is weak. Regulatory frameworks could require tech companies to assess and mitigate risks to democracy. For example, legislation might require platforms to allow vetted researchers to audit their algorithms for biases or political distortion, as in algorithmic transparency laws being discussed in the EU and elsewhere.

The key point is that *the democratic impact of technology is a result of design choices, which in turn reflect human beliefs and priorities.* Our coder survey finds a willingness among many tech workers to bring positive change, but also highlights the barriers they face. Overcoming those barriers could yield technologies that, by design, promote informed citizenship, protect rights, and foster constructive political communication. This would require both ideological and business-model shifts, but coders *could* implement it.

## *Democracy in the digital public sphere*

These findings exist within a digital public sphere increasingly split by content quality. The slop economy threatens democratic processes in multiple ways. Misinformation and afactual content can mislead voters by reducing the quality of democratic decision-making, erode trust in legitimate institutions, and exacerbate polarization. Moreover, these harms disproportionately affect those with fewer resources or digital savvy, thereby layering *information inequality on top of social inequality*. US consumers face a gap; the affluent/educated read paid sites like The New York Times or The Atlantic, while others subsist on free clickbait on Meta and TikTok. In much of the developing world, when a population comes online rapidly without strong media literacy or institutional supports, the slop flood can become *the* internet for users. Its consequences can include skewed perceptions of reality, difficulty in achieving consensus, and vulnerability of worse-off groups to exploitation by demagogues or scams.

Our findings support Read's (2024: np) warning that '*the slop tide threatens key functions of the web, clogging search results with nonsense and polluting the fragile information ecosystem*'. For democracies, this means citizens might need to exert significantly more effort to ensure that they are obtaining accurate information as opposed to false or irrelevant material (presuming that accuracy is their most important metric for information). In the US and UK, there are calls for platform regulation to ensure basic quality standards or transparency in algorithms. Yet, regulatory moves raise free speech issues and practical enforcement challenges, and may lead to a democratic electorate bifurcated into one group that engages in informed debate and another that is effectively living in a misinformation-rich bubble.

Addressing this divide likely requires technological, educational, and regulatory solutions. On the technology front, platforms could filter out the worst 'slop' content by down-ranking clickbait farms, labeling AI-generated content, and removing demonstrably false information. Search engines could re-program their algorithms to not surface obviously AI-scraped content in top results, or use public-interest algorithms: alternative recommendation systems that promote reliable information as Wikipedia has done for static pages.

Governments are grappling with how to hold platforms accountable without stifling free expression through mandating transparency reports on content quality metrics or algorithmic diversity in news feeds. Extreme versions of this debate include proposals to treat major social platforms as public utilities with obligations to serve the public good. A less intrusive approach is funding public service digital platforms. For instance, national broadcasters like the BBC could develop algorithms/feeds that compete with commercial platforms by offering personalized yet high-quality content streams free for users.

We also consider economic drivers. The slop economy exists because it's profitable: produce cheap content, gain clicks, earn ad revenue. If regulators or industry coalitions pressured ad networks to exclude known content farms, it could defund slop producers, as in efforts to demonetize extremist content (Ruiz, 2025). However, the slop economy is not just misinformation. Even content that is not outright false can be slop if it's contextless, trivial, or empty engagement that displaces substantive information. A democracy erodes not only when lies spread, but also when citizens' attention is consumed by irrelevance and sensationalism at the expense of civic knowledge. Thus, the goal should not only be to fight falsity, but to elevate substance. This is a harder challenge – one can't legislate people to read serious news after all. But societies have instituted norms around media for such purposes before, as when the U.S. Fairness Doctrine enforced balanced broadcasting. The guiding principle should be making reliable information ubiquitous and easy to consume, so that even those not actively seeking it will encounter it.

### *Bridging the gaps: toward an inclusive and ethical digital future*

We next discuss how coder beliefs and the slop economy intersect. There is emerging evidence, supported by our survey findings, of a reinforcing feedback loop between beliefs and digital ecosystems: the choices of tech creators influence the shape of the information environment, which in turn influences public beliefs and behaviors, which then feed back into what kinds of technologies are engaged most heavily with (Santos et al., 2021). This has a long evidence base in media studies, dating to the first mass newspapers in nineteenth century England (Jones, 2016). What is unclear today is how coders may influence, strengthen, or challenge this feedback loop. For the public, it appears to have negative tendencies. Tech firms design engagement-maximizing algorithms which create partisan, low-quality content ecosystems; this misinforms some public segments, who lose institutional trust as the algorithms build them unwitting echo chambers; companies respond by doubling down on engaging content to keep those users, and so on.

It raises a wicked problem: Even if tech companies pivot to more ethical design, if the public has already been primed to crave slop, they might reject or circumvent platforms that try to change their content to more 'boring' neutral quality. Conversely, even if users become more savvy and demand better info, they may fail if the vast majority of new apps run on slop engines to continue chasing engagement. Thus, an integrated approach is needed with stakeholders including tech workers, consumers, media, educators, and policymakers pushing together for a healthier digital public sphere. One promising concept is in 'democracy technologies' as a field analogous to 'health tech' or 'fintech', technology explicitly aimed at strengthening democratic society (Miklian & Hoelscher, 2017). This could include platforms for community deliberation, tools that facilitate transparency and fact-checking, algorithms that foster cross-cutting exposure to counter trends of slop and polarization. If coders with strong democratic values are empowered to innovate in this space, we might develop alternatives to the current social media paradigm, what some call 'digital public infrastructure.'

In bridging the gap between creators and consumers, diverse stakeholders can be incorporated into tech design via participatory design and co-creation processes. Participatory design has been championed in HCI to ensure technology meets social needs; applying it here could surface concerns that developers or executives might miss, like how a seemingly minor UI change could confuse less literate users or how a certain content recommendation pattern might fuel rumors. This kind of techno-solution should not overshadow the root causes, but it could be part of the interim measures to cope.

## Limitations and further research

We see three main limitations. First, our survey sample of coders, while novel, is not a perfectly representative or randomized sample of all developers. Thus, the results should be interpreted as exploratory. For example, the 'neutral/centrist' positions may be capturing Network State / Accelerationist adherents who call themselves non-partisan but have radical authoritarian and anti-democracy worldviews. Future research could aim for a more global survey of tech workers, possibly in partnership with professional associations, to validate these findings and delve deeper by examining differences between regions, or tracking changes over time as public debate on tech ethics evolves. There are also inherent challenges within a survey that examines potentially sensitive political and ethical issues: desirability bias may influence how candidates responded; or the survey may not have been able to fully capture the complex social problems related to socio-political values. Qualitative follow-up interviews with developers should also be seen as essential for theory building (Kim, 2025). This can enrich understandings of the dynamics we quantify here, such as in how exactly do tech workers negotiate ethical challenges, or what would enable them to say no more often.

Second, we treated the two threads (coder beliefs and digital divide) somewhat separately in our analysis and causal connections are not possible from this analysis alone; future work could aim to directly test these proposed connections. For example, one could investigate whether tech companies led by more socially conscious teams produce measurably different outcomes in the public's information environment. Or, scholars could examine specific technologies from design to user impact by documenting how decisions by engineers trickled to what users consumed. This could strengthen causality claims: we assert that coder worldviews influence products which influence society, but verifying that chain empirically requires multi-method research across disciplines. We encourage interdisciplinary collaborations (with e.g., computer scientists, sociologists, political scientists, media scholars) to more comprehensively tackle this.

Third, the evolving nature of technology means this research is a moving target. Research from even 2–3 years ago might be outdated by new developments (e.g., the explosion of ChatGPT-like tools in 2024), rewarding more rapid, continuous study models. One idea is a 'digital democracy observatory' that continuously monitors trends, akin to how some labs monitor social media. Policymakers and technologists would benefit from real-time insights as much as from academic papers. Therefore, we see this work less as part of a larger conversation and invite others to build on it, critique it, and test solutions in practice. We offer the full survey dataset on an open access basis for such purposes.

Technological innovation and democratic development are deeply intertwined in our contemporary world. We aimed to interrogate that connection that by examining the beliefs of those who create digital technologies and the way those technologies are experienced by citizens. While many tech creators are troubled by the democratic implications of their work, systemic pressures often prevent them from acting. Meanwhile, billions of users find themselves in a degraded online information environment that widens societal divides and weakens informed citizenship.

Democratizing the digital realm requires effort at both the production and consumption levels of technology. Democracy relies on shared truths, inclusive dialogue, and the agency of informed citizens. If the digital technologies that mediate so much of our public life work against those pillars, democracy itself is imperiled. Conversely, steering technology towards supporting transparency, reasoned debate, and equitable access to knowledge can revitalize democratic practice for the digital age. We acknowledge that implementing these ideas is a tall order, pushing against forces of profit, inertia, and user habits.. Our study aimed to contribute to this discussion by evidencing how these problems manifest at both the production and consumption ends of technology.

## Notes

1. We acknowledge extensive scholarship in the political economy of technology, capitalism and the sociopolitical outcomes and civic responses of these intersections (Couldry & Mejias, 2019; Wark, 2021; Sadowski, 2025).
2. For a discussion on the concept of a 'tech worker', see Niebler (2025).
3. See Vatillum (2019) for more on methodology, adapted by Miklian and Hoelscher (2017).

## Acknowledgments

The authors thank our survey participants, and are grateful to Alma Culén, Jasmin Neiss, Nicholas Stevens, John Katsos, Andreas Jungherr, Dan Mercea, Sophie Wade, and three anonymous reviewers for their suggestions and insights throughout the revision and editorial process.

## AI statement

Authors used ChatGPT for reference formatting purposes.

## Author contributions

CRediT: **Jason Miklian:** Conceptualization, Data curation, Formal analysis, Investigation, Methodology, Project administration, Resources, Validation, Visualization, Writing – original draft, Writing – review & editing; **Kristian Hoelscher:** Data curation, Formal analysis, Investigation, Methodology, Project administration, Resources, Writing – original draft, Writing – review & editing.

## Disclosure statement

No potential conflict of interest was reported by the author(s).

## Ethics statement

Ethical approach was aligned to guidance as given by Norway's National Research Ethics Committee (NREC), and adhered to NREC protocols for survey research of anonymized individuals per the National Committee for Research Ethics in the Social Sciences and the Humanities (NESH) for research participants. Informed consent was provided by RIWI as co-designed by the authors and RIWI as aligned with the NESH protocols, including informed consent and the responsibility to inform. The following text preceded the survey, and participants could opt-out at any time:

"This survey is being carried out by researchers at the University of Oslo to understand worldviews about technology and democracy. This survey is entirely anonymous and you will not be asked for any personal identifying information like your name or address. This survey is expected to take approximately 15 minutes. By continuing with the survey, you are indicating your willingness to participate in this research study and acknowledge that you are over the age of 18. To learn more, click on the "Privacy Policy" link below:"

## Funding

This work was supported by University of Oslo under the UiO: Democracy initiative, grant number UL105340100.

## Notes on contributors

*Jason Miklian*, Ph.D., is a Research Professor at the Center for Development and Environment, University of Oslo. Miklian has published over 60 academic and policy works on issues of crisis and responsible business, based on extensive fieldwork in Bangladesh, Colombia, India, and the DRC, among others. He serves on the United Nations Expert Panel on Business and Human Rights, and serves as an expert resource for various government knowledge banks in the US, UK, EU and Norway.

*Kristian Hoelscher* holds a PhD in Political Science from the University of Oslo, an MSc. in Population and Development Studies from the London School of Economics and Political Science, and a BA in Business and BSc (Hons.) in Psychology from the University of Queensland in Brisbane, Australia. His research interests lie at the intersection of cities, technology, politics, conflict and development in the Global South. He has worked in Sub-Saharan Africa, South Asia and Latin America, both quantitatively and qualitatively, and using interdisciplinary approaches.

## Appendix

**Appendix A: Survey Questionnaire**

*(The full questionnaire is redacted but available upon request)*